\documentclass[prx,aps,amssymb,twocolumn]{revtex4-1}
\usepackage{amsmath}
\usepackage{amssymb}
\usepackage{amsthm}
\usepackage{amsfonts}
\usepackage{listings}
\usepackage{tabularx}
\usepackage{makecell}
\usepackage{soul, color,xcolor}
\usepackage{enumerate}
\usepackage{latexsym}
\usepackage{color}
\usepackage{bm}
\usepackage{hyperref}
\hypersetup{
 pdfnewwindow=true, colorlinks=true,
 linkcolor=blue, anchorcolor=blue,
 citecolor=blue, filecolor=blue,
 menucolor=blue, urlcolor=blue}

\usepackage{psfrag}

\usepackage{bm}
\usepackage{graphicx}
\usepackage{subfigure}

\soulregister{\cite}7
\soulregister{\ref}7
\RequirePackage[normalem]{ulem} 
\RequirePackage{color}\definecolor{RED}{rgb}{1,0,0}\definecolor{BLUE}{rgb}{0,0,1} 

\input{epsf}

\newcommand{\nc}{\newcommand}
\nc{\fig}[1]{Fig. \ref{#1}}
\nc{\webirvsp}{\href{https://github.com/zjwang11/irvsp}{\texttt{IRVSP}} }
\nc{\webirtb}{\href{https://github.com/zjwang11/irvsp}{\texttt{ir2tb}} }
\nc{\webirpw}{\href{https://github.com/zjwang11/ir2pw}{\texttt{IR2PH}} }
\nc{\webchecktopmat}{\href{https://www.cryst.ehu.es/cryst/checktopologicalmagmat}{\texttt{Check Topological Mat}}}
\nc{\webposbr}{\href{https://github.com/zjwang11/UnconvMat/blob/master/src_pos2aBR.tar.gz}{\texttt{POS2ABR}} }
\nc{\webUnconvMat}{\href{http://tm.iphy.ac.cn/UnconvMat.html}{\texttt{UnconvMat}} }
\nc{\online}{\href{http://tm.iphy.ac.cn/UnconvMat.html}{online}}

\begin{document}

\tolerance 10000

\newcommand{\vk}{{\bf k}}

\draft

\title{Unconventional phonon spectra and obstructed edge phonon modes}


\author{Ruihan Zhang}
\affiliation{Beijing National Laboratory for Condensed Matter Physics,
and Institute of Physics, Chinese Academy of Sciences, Beijing 100190, China}
\affiliation{University of Chinese Academy of Sciences, Beijing 100049, China}

\author{Haohao Sheng}
\affiliation{Beijing National Laboratory for Condensed Matter Physics,
and Institute of Physics, Chinese Academy of Sciences, Beijing 100190, China}
\affiliation{University of Chinese Academy of Sciences, Beijing 100049, China}

\author{Junze Deng}
\affiliation{Beijing National Laboratory for Condensed Matter Physics,
and Institute of Physics, Chinese Academy of Sciences, Beijing 100190, China}
\affiliation{University of Chinese Academy of Sciences, Beijing 100049, China}

\author{Zhong Fang}
\affiliation{Beijing National Laboratory for Condensed Matter Physics,
and Institute of Physics, Chinese Academy of Sciences, Beijing 100190, China}
\affiliation{University of Chinese Academy of Sciences, Beijing 100049, China}

\author{Zhilong Yang}
\email{zhlyang@iphy.ac.cn}
\affiliation{Beijing National Laboratory for Condensed Matter Physics,
and Institute of Physics, Chinese Academy of Sciences, Beijing 100190, China}
\affiliation{University of Chinese Academy of Sciences, Beijing 100049, China}

\author{Zhijun Wang}
\email{wzj@iphy.ac.cn}
\affiliation{Beijing National Laboratory for Condensed Matter Physics,
and Institute of Physics, Chinese Academy of Sciences, Beijing 100190, China}
\affiliation{University of Chinese Academy of Sciences, Beijing 100049, China}


\begin{abstract}
Based on the elementary band representations (EBR), many topologically trivial materials are classified as unconventional ones (obstructed atomic limit), where the EBR decomposition for a set of electronic states is not consistent with atomic valence-electron band representations. In the work, we identify that the unconventional nature can also exist in phonon spectra, where the EBR decomposition for a set of well-separated phonon modes is not consistent with atomic vibration band representations (ABR). The unconventionality has two types: type I is on an empty site; and type II is on an atom site with non-atomic vibration orbitals. The unconventionality is described by the nonzero real-space invariant at the site. Our detailed calculations show that the black phosphorus (BP) has the type I unconventional phonon spectrum, while 1H-MoSe$_2$ has the type II one, although their electronic structures are also unconventional. Accordingly, the obstructed phonon modes are obtained for two types of unconventional phonon spectra.
\end{abstract}

\maketitle
\section{Introduction}
\begin{figure}
    \centering
     \includegraphics[width=0.98\linewidth]{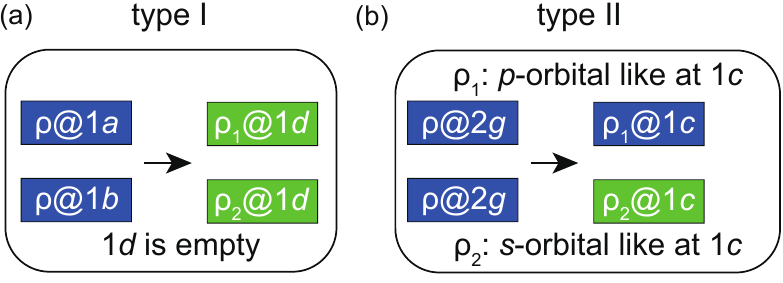}
    \caption{Two types of unconventionality. Atomic vibrations act as $p$ orbitals.
    \textbf{(a)} EBR at an empty site. 
    \textbf{(b)} EBR at an atom site with non-atomic vibration orbitals. 
    }
    \label{fig:type12.pdf}
\end{figure}

Recently, in topologically trivial materials, researchers have established a new class of unconventional materials (obstructed atomic limit)~\cite{TQC,PhysRevB.103.205133,PhysRevResearch.3.L012028,GaojcUnconvMat,li2111,xu2021,shaoresearch}, where the EBR decomposition is not consistent with atomic valence-electron band representations in the atomic limit. Although there is no topologically protected surface/edge state, obstructed electronic states usually emerge on their boundaries. Due to such boundary states, these materials have high chemical reactivity and various properties, such as low work function, strong hydrogen affinity, and electrocatalysis ~\cite{li2111,GaojcUnconvMat}. Recently, the unconventionality has two types: type I is on an empty site; and type II is on an atom site with non-atomic vibration orbitals, which is shown in \fig{fig:type12.pdf}. In the type I unconventional compound with an occupied EBR at an empty site, there is a filling anomaly in open-boundary conditions: a mismatch between the number of electrons in a symmetric geometry and the number of electrons required for charge neutrality~\cite{multipoleScience,PhysRevB.96.245115,PhysRevResearch.1.033074,xu2021fillingenforced}.  The type II unconventionality is found in phonon spectra, which has an EBR at an atom with non-atomic vibration orbitals~\cite{zhang2022large,Xu2022PhTop}.

In phonon spectra, the phonon bands split into some well-separated parts. For each part, one can identify unconventionality and topology~\cite{zhang2022large,Xu2022PhTop}. In this work, we defined an unconventional phonon spectrum, where the EBR decomposition of the phonon modes is not consistent with atomic vibration band  representations (ABR; not a sum of ABR)~\cite{GaojcUnconvMat}. As a result, the obstructed phonon band can emerge in the unconventional frequency gap, as found in element Se/Te~\cite{zhang2022large}. Based on first-principles calculations and computed phonon spectra, we find that BP and 1H-MoSe$_2$ have  both unconventional phonon spectra and electronic band structures. The obstructed phonon modes are obtained on the edges.

\section{Results and discussions}

\subsection{Unconventional electronic band structure in black phosphorus}

\begin{figure}
    \centering
     \includegraphics[width=0.98\linewidth]{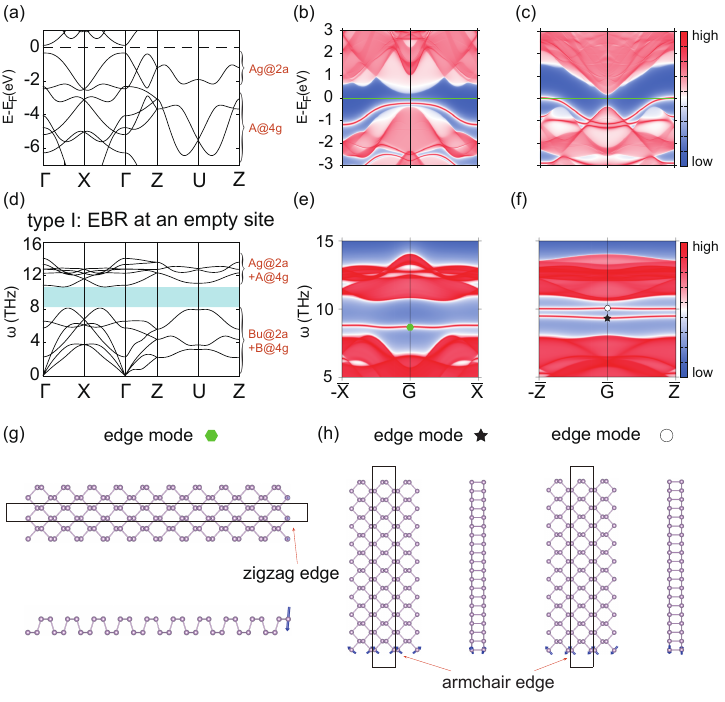}
    \caption{The 2D bulk states, band representations, and obstructed states of BP monolayer.
    \textbf{(a-c)} Electronic band structure with unconventional EBRs [$A@4g+A_g@2a$], obstructed edge states on the zigzag  and armchair edges. \textbf{(d-f)} Phonon spectrum with unconventional EBRs [$B@4g+B_u@2a$ (lower six bands) and $A@4g + A_g@2a$ (higher six bands)], obstructed phonon modes on the zigzag  and armchair edges. \textbf{(g,h)} Vibrations of obstructed phonon edge modes.
    }
    \label{fig:bp.pdf}
\end{figure}

We first review the unconventional nature in the electronic band structure.
The space group (SG) of BP monolayer is	$Pmna$ (SG \#53), and the P atoms are located at the $4h$ Wyckoff position. Thus, the atomic valence-electron band representations of P-$p$ orbitals are $A^{\prime}@4h$ and $A^{\prime\prime}@4h$ (assigned by \webposbr code). The band structure of BP is shown in 
 \fig{fig:bp.pdf}(a). The calculation details are given in Appendix A. The EBR decomposition of occupied bands shows that the six occupied $p$-orbital bands belong to $A@4g + A_g @ 2a$ EBRs (four $s$-orbital bands belong to EBR $A^{\prime}@4h$).  
Obviously, the $4g$ and $2a$ Wyckoff positions are not filled by any atoms, which are in the middle of the P--P bonds. The real-space invariant (RSI) for the empty 4g (2a) site is computed to be $\delta_{1}@4g\equiv-m(A)+m(B)=-1$ $[ \ \delta_{1}@2a \equiv - m(A_{g})+m(A_{u})- m(B_{g})+ m(B_{u})=-1 \ ]$, where $m(\rho$) denotes the number of $\rho$@4g, indicating the unconventional nature of the obstructed atomic insulator (OAI). Due to its unconventional electronic structure, there exist obstructed edge states in zigzag and armchair edges, as presented in Figs. \ref{fig:bp.pdf}(b) and \ref{fig:bp.pdf}(c). 

\begin{table}[t!]
\centering
\caption{The atomic vibration phonon band representations of BP monolayer (SG 53).}
\label{tab:bp_aBR}
\begin{tabular*}{\linewidth}{p{1.1cm}|p{1.1cm}|p{1.3cm}|p{2.2cm}|p{2.2cm}}
    \hline
    \hline
    \makecell[c]{Atom} & \makecell[c]{WP($q$)} & \makecell[c]{Symm.} & \makecell[c]{Vibrations ($\rho$)} & \makecell[c]{ABRs($\rho@q$)} \\
    \hline
      \makecell[c]{P}  & \makecell[c]{$4h$} & \makecell[c]{$m$} & \makecell[c]{$p_{x}$ :\: $A^{\prime}$}  & \makecell[c]{$A^{\prime}@4h$}  \\
         &      &     & \makecell[c]{$p_{y}$ :\: $A^{\prime}$}  & \makecell[c]{$A^{\prime}@4h$}  \\
         &      &     & \makecell[c]{$p_{z}$ :\: $A^{\prime\prime}$} & \makecell[c]{$A^{\prime\prime}@4h$}  \\
    \hline
    \hline
\end{tabular*}

\end{table}

\subsection{Type I unconventional phonon spectra in BP}

Then, we calculated its phonon spectrum, as shown in \fig{fig:bp.pdf}(d). In the phonon spectrum, the twelve phonon bands split into two parts with a large frequency gap. Our calculations show that the lower  six bands belong to $B@4g + B_u@2a$ (\webirpw code). The RSI for the empty 4g (2a) site (lowest six bands) is computed to be $\delta_{1}@4g = 1  (\delta_{1}@2a = 1)$. On the other hand, as the atomic vibrations act with the $p$-orbital symmetry, the ABRs are listed in Table \ref{tab:bp_aBR}. The EBR decomposition implies that the well-separated phonon bands have type I unconventionality with an EBR at an empty site, and the obstructed phonon edge mode can emerge. One zigzag edge mode and two armchair edge modes are obtained in the open-boundary calculations, as shown explicitly in Figs. \ref{fig:bp.pdf}(e) and \ref{fig:bp.pdf}(f). Their phonon vibrations are shown in Figs. \ref{fig:bp.pdf}(g) and \ref{fig:bp.pdf}(h). It can be seen that these phonon modes are primarily composed of the vibrations of boundary atoms. The unconventionality of its phonon spectrum can explain the experimental observation of edge phonon modes in BP monolayers~\cite{BP1,BP2}.

\begin{figure}
    \centering
    \includegraphics[width=0.98\linewidth]{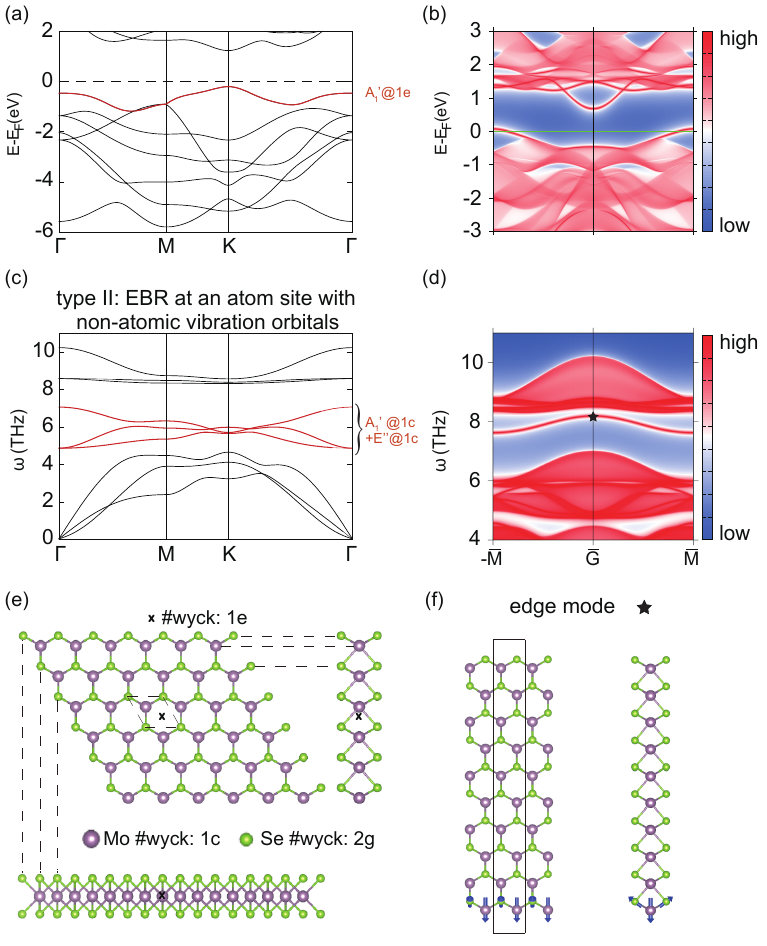}
    \caption{The 2D bulk states, band representations, and obstructed states of 1H-$\mathrm{MoSe_{2}}$.
    \textbf{(a,b)} Electronic band structure with an unconventional EBR [$A_1^{\prime}@1e$ (red band)], obstructed edge states on the zigzag edge.
    \textbf{(c,d)} Phonon spectrum with unconventional EBRs [$A_{1}^{\prime}@1c+E^{\prime\prime}@1c$ (three red bands)], obstructed phonon modes on the zigzag edge.
    \textbf{(e,f)} Crystal structure and vibrations of obstructed phonon edge mode.
    }
    \label{fig:mose.pdf}
\end{figure}

\begin{table}[t!]
    \centering
    \caption{
    The atomic vibration phonon band representations of 1H-MoSe$_2$ (SG 187).
    }
    \label{tab:mose_aBR}
    \begin{tabular*}{\linewidth}{p{1.1cm}|p{1.1cm}|p{1.3cm}|p{2.2cm}|p{2.2cm}}

    \hline
    \hline
        \makecell[c]{Atom} & \makecell[c]{WP($q$)} & \makecell[c]{Symm.} & \makecell[c]{Vibrations ($\rho$)} & \makecell[c]{ABRs($\rho@q$)} \\
        \hline
                  \makecell[c]{Mo} & \makecell[c]{$1c$} & \makecell[c]{$-62m$} & \makecell[c]{$p_x, p_{y}$ :\: $E^{\prime}$}  & \makecell[c]{$E^{\prime}@1c$}  \\
             &      &        & \makecell[c]{$p_{z}$ :\: $A_{2}^{\prime\prime}$} & \makecell[c]{$A_{2}^{\prime\prime}@1c$}  \\
        \hline
          \makecell[c]{Se} & \makecell[c]{$2g$} & \makecell[c]{$3m$} & \makecell[c]{$p_{x}, p_y$ :\: $E$ } & \makecell[c]{$E@2g$}  \\
             &      &       & \makecell[c]{$p_{z}$ :\: $A_{1}$ }& \makecell[c]{$A_{1}@2g$ } \\
        \hline
        \hline
    \end{tabular*}
\end{table}

\subsection{Type II unconventional phonon spectra in 1H-phase MoSe$_2$}
In this section, we focus on the monolayer of the 2H phase (1H-phase), our calculations show that the 1H-phase MoSe$_2$ has obstructed electronic and phonon band structures. The crystal of 1H-MoSe$_2$ is shown in \fig{fig:mose.pdf}(e). It possesses a hexagonal structure with a SG of $P6_3/mmc$ (SG \#187), where the Mo atoms plane is sandwiched by two Se atom planes, which sit in the $1c$ and $2g$ Wyckoff positions, respectively.
The bulk band structure and the EBR decomposition of 1H-MoSe$_2$ are shown in \fig{fig:mose.pdf}(a). It can be seen that the highest valence band (red-colored) belongs to $A_{1}^{\prime}@1e$ ($1e$ is empty). The RSI for the empty 1e site is computed to be $\delta_{1}@1e \equiv m(A_{1}^{\prime}) + m(A_{1}^{\prime\prime}) - m(A_{2}^{\prime\prime}) - m(A_{2}^{\prime}) - m(E^{\prime}) + m(E^{\prime\prime}) = 1 $. As a result, the obstructed states are obtained in open-boundary directions, as shown in \fig{fig:mose.pdf}(b).
The obstructed band structures of other 1H-phase transition metal dichalcogenides (TMDs, 1H-M$X_2$) can be found in Appendix B.

In its phonon spectrum, the nine phonon bands split into three parts, as shown in \fig{fig:mose.pdf}(c). Using \webirpw code, we find that the three parts belong to  $A_{2}^{\prime\prime}@1c+E^{\prime}@1c,A_{1}^{\prime}@1c+E^{\prime\prime}@1c$ (red-colored), and $ A_{2}^{\prime\prime}@1c+E^{\prime}@1c$, respectively. The ABRs for the phonon bands are obtained in Table \ref{tab:mose_aBR}. The RSI for the occupied 1c site (three red bands) is computed to be $\delta_{1}@1c \equiv m(A_{1}^{\prime}) + m(A_{1}^{\prime\prime}) - m(A_{2}^{\prime\prime}) - m(A_{2}^{\prime}) - m(E^{\prime}) + m(E^{\prime\prime}) = 2 $. Although the middle three phonon bands representations are centered on the $1c$ Wyckoff position of Mo atoms, its vibration character is not $p$-orbital like. On the other hand, the vibration modes of Mo ($@1c$) only support three phonon bands of $A_{2}^{\prime\prime}@1c+E^{\prime}@1c$, corresponding to the lowest part (three acoustic phonon bands). 
Therefore, the middle three phonon bands should not be of the atomic vibration limit, and the unconventionality of 1H-MoSe$_2$ phonon spectrum is type II with an EBR at an atom site with non-atomic vibration orbitals.
Accordingly, the obstructed phonon mode is obtained in the frequency gap, as shown in \fig{fig:mose.pdf}(d). The vibration of obstructed edge phonon mode is shown in \fig{fig:mose.pdf}(f).

\subsection{Janus 1H-phase MoSSe}

\begin{figure}
    \centering
    \includegraphics[width=0.98\linewidth]{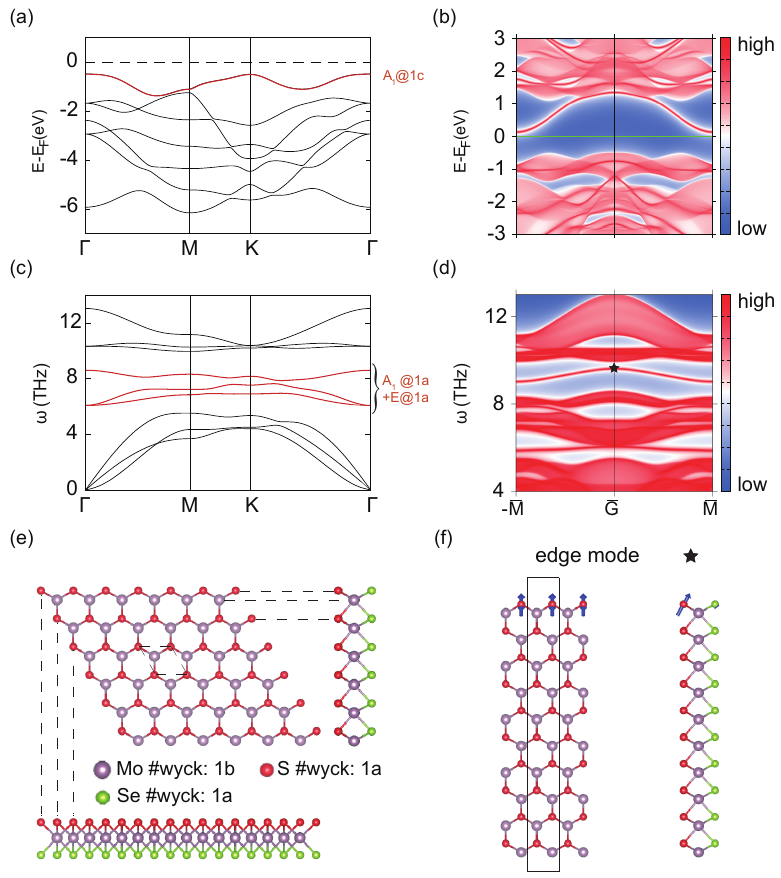}
    \caption{The 2D bulk states, band representations, and obstructed states of Janus 1H-$\mathrm{MoSSe}$.
\textbf{(a,b)} Electronic band structure with an unconventional EBR [$A_{1}@1c$ (red band)], obstructed edge states on the zigzag edge.
    \textbf{(c,d)} Phonon spectrum, obstructed phonon modes on the zigzag edge.
    \textbf{(e,f)} Crystal structure and vibrations of obstructed phonon edge mode.
    }
    \label{fig:mosse.pdf}
\end{figure}

The Janus 1H-phase MoSSe is obtained by replacing Se with S in the top layer of 1H-phase MoSe$_2$. The crystal structure is shown
in \fig{fig:mosse.pdf}(e), and the symmetry becomes $P3m1$ (SG \#156) due to the atom substitution. The $1c/1e$ and $2g$ sites of SG 187 become $1b/1c$ and $1a$ sites of SG 156, respectively. As a result, the highest band in the electronic band structure becomes $A_{1}@1c$ EBR in SG 156, as shown in \fig{fig:mosse.pdf}(a).
The RSI for the empty 1c site is computed to be $\delta_{1}@1c \equiv - m(A_{1}) - m(A_{2}) + m(E)  = -1$. The $1c$ is empty and the unconventionality of electronic structure is type I. The obstructed electronic states are shown in \fig{fig:mosse.pdf}(b).

As the EBR analysis strongly depends on the symmetry eigenvalues and irreps, it may not work for some lower-symmetry cases. As we find the unconventionality analysis in the phonon spectrum of Janus 1H-phase MoSSe fails due to the lack of the mirror $z$ symmetry. However, the unconventionality of the offset  phonon-mode centers may still exist. We directly compute the edge spectra of the phonon structure, as shown in \fig{fig:mosse.pdf}(d), and the obstructed edge phonon vibration is shown in \fig{fig:mosse.pdf}(f).

\section{Discussion}
The unconventionality widely exists in the electronic and phononic systems, where the EBR decomposition of the band structure is not consistent with atomic valence-electron/vibration band representations (not a sum of ABR). 
Two types of unconventionality are found in phonon spectra: type I is on an empty site; and type II is on an atom site with non-atomic vibration orbitals.
We find that the unconventional BP phonon band structure is type I, while the unconventional MoSe$_2$ is  type II.
More importantly, both electronic and phonon states are unconventional in these materials. Due to the coexistence of obstructed electronic states and phonon modes, the electron-phonon coupling strength may be enhanced and may generate superconductivity at the boundary.

\ \\
\noindent \textbf{Acknowledgments}
This work was supported by the National Natural Science Foundation of China (Grants No. 11974395, No. 12188101), the Strategic Priority Research Program of Chinese Academy of Sciences (Grant No. XDB33000000), National Key R\&D Program of Chain (Grant No. 2022YFA1403800), and the Center for Materials Genome.

\bibliography{tps}



\section*{Appendix}

\begin{figure}[b!]
    \centering
    \includegraphics[width=0.98\linewidth]{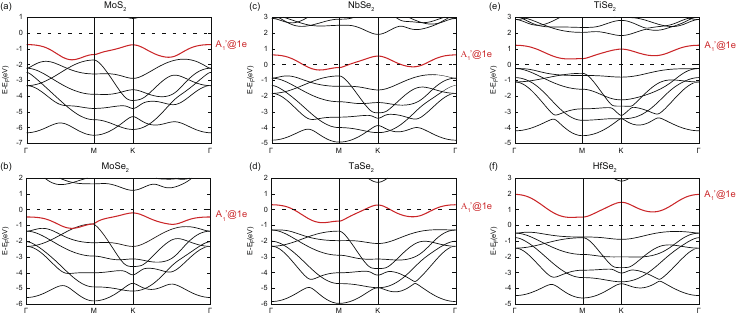}
    \caption{ The band structures of
    \textbf{(a)} $\mathrm{MoS_2}$,
    \textbf{(b)} $\mathrm{MoSe_2}$,
    \textbf{(c)} $\mathrm{NbSe_2}$,
    \textbf{(d)} $\mathrm{TaSe_2}$,
    \textbf{(e)} $\mathrm{TiSe_2}$,
    \textbf{(f)} $\mathrm{HfSe_2}$.
    The target band (red colored) belongs to $A_{1}'@1e$, being the obstructed atomic limit and of type I unconventionality.}
    \label{fig:br.pdf}
\end{figure}

\subsection{Methodology}
We performed the first-principles calculations based on the density functional theory (DFT) implemented in the Quantum ESPRESSO package~\cite{Giannozzi_2009, Giannozzi_2017}, using the projector augmented-wave (PAW) method~\cite{PhysRevB.50.17953PAW1, PhysRevB.59.1758PAW2}. 
The electrons are described using the generalized gradient approximation (GGA) with the exchange-correlation functional of Perdew, Burke and Ernzerhof (PBE)~\cite{GGA-PBE1996} for the exchange-correlation functional. The thickness of the vacuum layer was set to larger than $15$ \AA. The lattice dynamical properties are described using the GGA with exchange-correlation functional of PBE for the exchange-correlation functional. The plane-wave cutoff used for BP, MoSe$_2$, and MoSSe calculations is 70, 60, and 90 Ry, respectively. The structures are relaxed until the forces reached a convergence threshold of $1.0\times{10}^{-7}$ Ry/Bohr. For the edges states calculation of 1D ribbons, we set the Wannierized ribbon width to 40 cell widths. The edge structure is directly cut from the 2D lattice without relaxation, in order to illustrate the unconventionality of the 2D materials. The irreducible representations of the phonon modes were obtained by the program \webirpw~\cite{zhang2022large,GaojcIRVSP}. The EBR/ABR decomposition of the phonon band structure is done on the  \webUnconvMat website~\cite{GaojcUnconvMat} with the obtained  \texttt{tqc.data}.

\subsection{Band structures of 1H-M$X_2$}
The empty-site EBR is very robust in the series of the 1H-M$X_2$. The M and $X$ are located at 1c and 2g Wyckoff positions, while the red-colored band belongs to $A_{1}'@1e$ (1e is an empty site), being the obstructed atomic limit and of type I unconventionality. One can change the number of electrons with different M atoms. In Mo$X_2$, the empty-site EBR is fully occupied, as shown in Figs.~\ref{fig:br.pdf}(a) and \ref{fig:br.pdf}(b). In Nb/Ta$X_2$, this EBR is half occupied [ Figs.~\ref{fig:br.pdf}(c) and \ref{fig:br.pdf}(d)]. In Ti/Hf$X_2$, this EBR is fully unoccupied. However it is still well isolated [ Figs.~\ref{fig:br.pdf}(e) and \ref{fig:br.pdf}(f)].


%

\end{document}